Optical and structural properties of Si doped β-Ga$_2$O$_3$ (010) thin films homoepitaxially grown by halide vapor phase epitaxy

Bahadir Kucukgok[1], David J. Mandia[2], Jacob H. Leach[3], Keith R. Evans[3], Jeffrey A. Eastman[2], Hua Zhou[4], John Hryn[1], Jeffrey W. Elam[1], and Angel Yanguas-Gil[*1]

[1]Applied Materials Division, Argonne National Laboratory, Lemont, Illinois, United States;

[2]Materials Science Division, Argonne National Laboratory, Lemont, Illinois, United States;

[3]Kyma Technologies, Inc., Raleigh, North Carolina, United States;

[4]X-ray Science Division, Argonne National Laboratory, Lemont, Illinois, United States.

## Abstract

We report the optical, electrical, and structural properties of Si doped β-Ga$_2$O$_3$ films grown on (010)-oriented β-Ga$_2$O$_3$ substrate via HVPE. Our results show that, despite growth rates that are more than one order of magnitude faster than MOCVD, films with mobility values of up to 95cm$^2$V$^{-1}$s$^{-1}$ at a carrier concentration of $1.32\times10^{17}$cm$^{-3}$ can be achieved using this technique, with all Si-doped samples showing n-type behavior with carrier concentrations in the range of $10^{17}$ to $10^{19}$cm$^{-3}$. All samples showed similar room temperature photoluminescence, with only the samples with the lowest carrier concentration showing the presence of a blue luminescence, and the Raman spectra exhibiting only phonon modes that belong to β-phase Ga$_2$O$_3$, indicating that the Ga$_2$O$_3$ films are phase pure and of high crystal quality. We further evaluated the epitaxial quality of the films by carrying out grazing incidence X-ray scattering measurements, which allowed us to discriminate the bulk and film contributions. Finally, MOS capacitors were fabricated using ALD HfO$_2$ to perform C-V measurements. The carrier concentration and dielectric values extracted from the C-V characteristics are in good agreement with Hall probe measurements. These results indicate that HVPE has a strong potential to yield device-quality β-Ga$_2$O$_3$ films that can be utilized to develop vertical devices for high-power electronics applications.

1. Introduction

Gallium oxide (Ga$_2$O$_3$) is an emerging ultrawide-bandgap semiconductor that could enable future generations of high-power electronic and ultraviolet (UV) optoelectronic devices. Its advantages include excellent chemical and thermal stability, high breakdown electric field (8MV/cm)[1]; i.e., almost three times larger than that of either SiC or GaN[2], large critical electric field (~5.2MV/cm)[3], and large bandgap (4.5-4.9eV)[4,5], which makes it transparent from the visible into the UV wavelength region and allows high-voltage and high temperature operation[6]. Furthermore, Baliga's figure of merit for the β-phase of Ga$_2$O$_3$ (β-Ga$_2$O$_3$) is several times larger than that of SiC or GaN due to its large bulk electron mobility (300cm$^2$V$^{-1}$s$^{-1}$)[7,8], which also results in higher breakdown voltages (V$_{br}$) and lower loss (i.e., higher efficiency)[9] than those of SiC and GaN-based power devices. Consequently, these favorable attributes of Ga$_2$O$_3$ make it an ideal candidate for applications in electronics such as metal-oxide-semiconductor field effect transistors (MOSFETs)[3], field-effect transistors (FET)[10], and Schottky barrier diodes (SBDs)[11] and optoelectronics such as solar-blind photodetectors[12]. However, in order to enable high-performance power devices, the ability to grow high-quality bulk single crystal substrates and thick homoepitaxial films with accurate control of dopant concentration and residual carrier concentration below $10^{16}$cm$^{-3}$ [13] are two key requirements.

Ga$_2$O$_3$ has five known phases (α, β, ε, γ, and δ) Among these, β-Ga$_2$O$_3$ is the most thermodynamically stable[14], while the other phases are metastable. Tin (Sn), germanium (Ge), and silicon (Si) have been successfully utilized as n-type dopants in β-Ga$_2$O$_3$[4,15,16] to obtain a wide-range of carrier concentrations from $10^{15}$ to as high as $10^{20}$cm$^{-3}$.[17] Notably, high-quality, large-area bulk β-Ga$_2$O$_3$ single crystal substrates have been synthesized by low cost and scalable melt-based growth techniques such as edge-defined film-fed growth (EFG)[18,19], floating zone (FZ)[20-23], and Czochralski[24-27], which enable β-Ga$_2$O$_3$ to be a superior candidate for next-generation high-power electronic applications. On the other hand, different growth methods, including molecular-beam epitaxy (MBE)[9,28-31], pulsed laser deposition (PLD)[32], mist chemical vapor deposition (mist-CVD)[33], metal-organic chemical vapor deposition (MOCVD)[34-36], low-pressure chemical vapor deposition (LPCVD)[14,37,38], and hydride vapor phase epitaxy (HVPE)[6,13,39,40] have been used to grow of β-Ga$_2$O$_3$ films. Nevertheless, growth techniques such as MOCVD and MBE result in low growth rates (<<1µm/h)[14] due to challenges in the growth kinetics.

---

[*] Email: ayg@anl.gov

In contrast, HVPE, an epitaxial growth technique, has significant advantages including rapid growth rates e.g., ~5µm/h on β-Ga$_2$O$_3$ (001)[13], doping control over a wide range, and higher throughput and lower-cost production. In this study, we have carried out room temperature (RT) characterization of the optical, structural, and electrical properties of Si doped β-Ga$_2$O$_3$ films grown by HVPE on semi-insulating Fe doped and bulk β-Ga$_2$O$_3$ substrates with unintentionally doped (UID) Ga$_2$O$_3$ buffer layers. A key motivation for our work is the fact that research on the electronic properties of on β-Ga$_2$O$_3$ has focused primarily on bulk β-Ga$_2$O$_3$ single crystal materials[41] with comparatively fewer reports about the electrical characterization of homoepitaxial β-Ga$_2$O$_3$ films, especially films grown by HVPE[42]. We have also chosen to study films grown with (010) orientation of β-Ga$_2$O$_3$, which has thus far been explored to a lesser extent than the (001) orientation.

## 2. Experimental

β-Ga$_2$O$_3$ films were grown by HVPE on (010)-oriented, semi-insulating bulk β-Ga$_2$O$_3$ single crystal substrates at Kyma Technologies Inc. The details of the HVPE growth parameters used to grow the bulk single crystal substrate and the thin film are explained in Ref 43. Si-doped Ga$_2$O$_3$ films were grown on a UID Ga$_2$O$_3$ buffer layer, with the total thickness of the grown stack in the 0.6-1.07µm range. Room temperature (RT) photoluminescence (PL) properties were studied using a Horiba Jobin Yvon Nanolog Spectrofluorometer equipped with a 450 Watt xenon lamp using a 260nm (4.76eV) excitation source and a 295nm low pass filter. The spectra were recorded with integration time and excitation grating of 1s and 1200 grooves/mm×330nm, respectively. A Renishaw inVia Raman Microscope equipped with 50-X focusing/collecting optic was used to carry out Raman spectroscopy measurements in backscattering geometry at RT. The Raman spectrometer employed a 532nm diode laser and a 1800 line/mm grating. Grazing incidence X-ray diffraction (GIXRD) measurements on non-specular out-of-plane Bragg peaks were performed on beamline 12ID-D of the Advanced Photon Source (APS), at Argonne National Laboratory. The incident X-ray beam hits the sample at a small angle (<5°) near the critical angle of total external X-ray reflection, and the X-ray energy was 20keV.

For the electrical characterization of the samples, a Cr (20nm)/Au(150nm) metal stack was deposited on a β-Ga$_2$O$_3$ thin film through a shadow mask using RF sputtering followed by 5min thermal annealing at ~475°C in a N$_2$ atmosphere to create Ohmic contacts. Hall effect measurements at RT using the van der Pauw method with a magnetic field of 0.51T were then conducted using an Ecopia HMS-3000 Hall Measurement System to determine the carrier concentration (n) and electron mobility (µ) in the thin films. HfO$_2$ films (2nm thickness) were deposited by thermal atomic layer deposition (ALD) on 1-inch-diameter, unintentionally doped bulk β-Ga$_2$O$_3$ (010) single crystal substrates (Kyma Inc.) grown by HVPE. The HfO$_2$ ALD was performed at 200°C in a custom viscous flow ALD reactor[44] using tetrakis(dimethylamido)hafnium (IV) (TDMAH) and H$_2$O. Subsequently, circular Au (150nm) topside contacts with a diameter of 650µm and a large area blanket Cr (20nm)/Au (150nm) backside contact were deposited by RF sputtering to facilitate capacitance-voltage (C-V) measurements using an Agilent E4980A precision LCR meter.

## 3. Results and Discussion

RT PL spectra of all the samples described in Table I are displayed in Fig. 1. There are some common features in the spectra of all the samples: as the PL excitation source used in this work is 4.76eV (260nm), no near-band-edge emission (NBE) was detected. Instead, we observe a strong broad-band UV luminescence that could be deconvoluted into three Gaussian components. The details of the strong-broad-band UV emission and the fitted data are illustrated in Table I. The consensus in the literature thus far is that this UV band is not associated with impurities, dopants, or growth conditions, but that instead it can be attributed to an intrinsic transition resulting from self-trapped excitations[45-48].

Of the four samples, only the sample with the lowest carrier concentration ($n=1.32\times10^{17}cm^{-3}$) includes an emission peak in the blue region of the spectrum (i.e. below 3.0eV). This blue luminescent band is caused mainly by donor-acceptor-pair (DAP) transitions including deep donors and acceptors. Possible donors and acceptors are oxygen vacancies ($V_o$) and interstitial Ga ($Ga_i$) and Ga vacancies ($V_{Ga}$) and $V_o$-$V_{Ga}$ complexes, respectively[45,46,48,49]. According to Binet and Gourier[49], blue luminescence results from an electron in a donor vacancy that is captured by an excited hole on an acceptor to generate a trapped exciton. The underlying mechanism of the blue luminescence intensity is related to the formation energy of $V_o$ donors (FE $V_o$), which is closely related to the Fermi level ($E_F$) position. As $E_F$ moves through the conduction band due to an increase in the carrier concentration, the FE $V_o$ rises[45]. Therefore, higher carrier concentration leads to a rapid reduction of $V_o$ donors and blue luminescence intensity. Consequently, samples that have higher carrier concentration do not exhibit the blue luminescence band (Fig. 1).

The peak positions of the broad UV emission are relatively similar; i.e., 3.34-3.35eV except for the sample having a carrier concentration of $1.32\times10^{17}cm^{-3}$, which is redshifted to 3.28eV. This redshift is due to the effect of low carrier concentration on the luminescence[50], which could also result in less lattice degeneration and carrier relaxation through the deep-gap-states before recombination occurs[21]. In the case of peaks centered at higher energies, as mentioned previously, the conduction band is starting to be occupied by electrons while the carrier concentration is increasing. As a result, $E_F$ rises and the PL peak shifts to higher energy since more electrons are present in the conduction band for the recombination process[51]. Moreover, a relatively less intense broadband peak centered in the range of 1.74-1.76eV (red emission) is also observed, which is related to the variation of bandgap structure of β-$Ga_2O_3$[52]. Red emission may originate from the intrinsic properties or recombination of electrons trapped on a donor owing to $V_o$ and a hole trapped on an acceptor due to nitrogen that possibly replaces the oxygen atoms in the lattice during the growth process[52-55]. The UV PL intensity showed no correlation with carrier concentration.

Table I. Information about PL emission bands derived from Gaussian fits of Fig. 1.

| Sample | Parameters | PL Energy (eV) | PL Linewidth (meV) |
|---|---|---|---|
| Si:$Ga_2O_3$ epi on Fe:$Ga_2O_3$ (010) ($n=1.02\times10^{19}cm^{-3}$) | Cumulative Fit Peak | 3.34 | 693 |
| | Gaussian 1 (UV) | 3.15 | 583 |
| | Gaussian 2 (UV) | 3.46 | 525 |
| | Gaussian 3 (UV) | 3.81 | 306 |
| Si:$Ga_2O_3$ epi on Fe:$Ga_2O_3$ (010) ($n=7.92\times10^{18}cm^{-3}$) | Cumulative Fit Peak | 3.35 | 699 |
| | Gaussian 1 (UV) | 3.13 | 568 |
| | Gaussian 2 (UV) | 3.45 | 529 |
| | Gaussian 3 (UV) | 3.78 | 344 |
| Si:$Ga_2O_3$ epi on $Ga_2O_3$ (010) ($n=4.04\times10^{17}cm^{-3}$) | Cumulative Fit Peak | 3.35 | 720 |
| | Gaussian 1 (UV) | 3.11 | 584 |
| | Gaussian 2 (UV) | 3.43 | 535 |
| | Gaussian 3 (UV) | 3.73 | 385 |
| Si:$Ga_2O_3$ epi on $Ga_2O_3$ (010) ($n=1.32\times10^{17}cm^{-3}$) | Cumulative Fit Peak | 3.28 | 672 |
| | Gaussian 1 (Blue) | 2.83 | 357 |
| | Gaussian 2 (UV) | 3.19 | 442 |

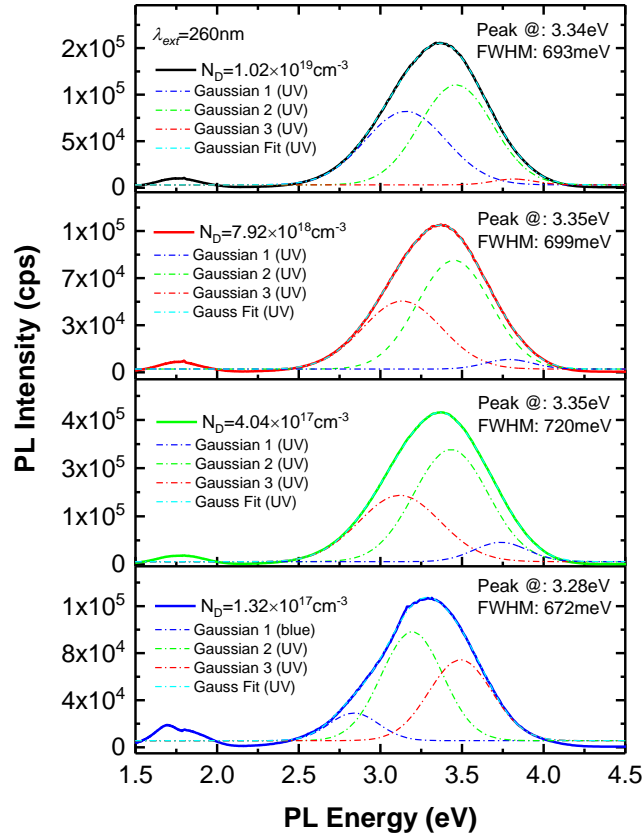

FIG. 1. RT PL spectra of Si doped β-Ga$_2$O$_3$ epitaxial layers grown on semi-insulating Fe doped β-Ga$_2$O$_3$ (010) (n>10$^{18}$cm$^{-3}$) and β-Ga$_2$O$_3$ (010) (n<10$^{18}$cm$^{-3}$) substrates. The dotted lines represent the fitted Gaussian curves.

We characterized the electrical properties of our β-Ga$_2$O$_3$ films grown on native substrates using both Hall probe and C-V measurements. In Fig. 2, we show the electron mobility versus the carrier concentration for homoepitaxial layers of Si doped β-Ga$_2$O$_3$ grown on Fe doped (n>10$^{18}$cm$^{-3}$) and undoped (n<10$^{18}$cm$^{-3}$) native substrates, as well as a comparison with literature values. Utilizing Si as a dopant, all samples exhibited n-type behavior with carrier concentrations between $1.32\times10^{17}$ and $1.02\times10^{19}$cm$^{-3}$. The highest electron mobility achieved was 95cm$^2$V$^{-1}$s$^{-1}$ at a carrier concentration of $1.32\times10^{17}$cm$^{-3}$. Electron mobility shows a decreasing trend with the increasing carrier concentration except for one sample, and the value of mobility varies from 67 to 95cm$^2$v$^{-1}$s$^{-1}$. This mobility behavior suggests that scattering at ionized impurities dominates when carrier concentrations are higher than 10$^{18}$cm$^{-3}$ [56-59]. In contrast, according to both experimental and theoretical studies of scattering mechanisms in β-Ga$_2$O$_3$[59-61], polar optical phonon scattering is the dominant mechanism for restricting the electron mobility to <200cm$^2$V$^{-1}$s$^{-1}$ at RT for carrier concentrations lower than ~2×10$^{18}$cm$^{-3}$ [56]. Note that the measured electron mobility of the β-Ga$_2$O$_3$ films is lower than the theoretically predicted values[42,62] due to phonon scattering and ionized impurity scattering, the major mechanisms limiting electron mobility in β-Ga$_2$O$_3$.

The sample with a carrier concentration value of $4.04\times10^{17}$cm$^{-3}$ shows the highest PL FWHM value of 720meV (Table I), indicating high crystal deterioration in the sample. As a result, it exhibits the lowest mobility value of 67cm$^2$V$^{-1}$s$^{-1}$ (see Fig. 1). We attribute this effect to a deterioration of the crystalline quality of the films with Si doping, which results in defect formation in the films and increased carrier scattering.

For comparison, the reported RT electron mobility of Si doped β-Ga$_2$O$_3$ films grown on semi-insulating Fe doped β-Ga$_2$O$_3$ and undoped β-Ga$_2$O$_3$ substrates with different crystal orientation by various growth methods are also shown (open symbols) in Fig. 2. For carrier concentration <10$^{18}$cm$^{-3}$, the mobility values of Ref. 63 and Ref. 6 are higher than those of this work. In contrast, our mobility values are larger than the corresponding literature values at carrier concentrations ≥10$^{18}$cm$^{-3}$. Our results validate that Si is an active n-type dopant for β-Ga$_2$O$_3$ films and imparts higher carrier concentrations with less reduction in electron mobility compared to tin (Sn) and germanium (Ge) dopants[4,15,29,30,41,58,63-66], setting a new standard for HVPE-grown films.

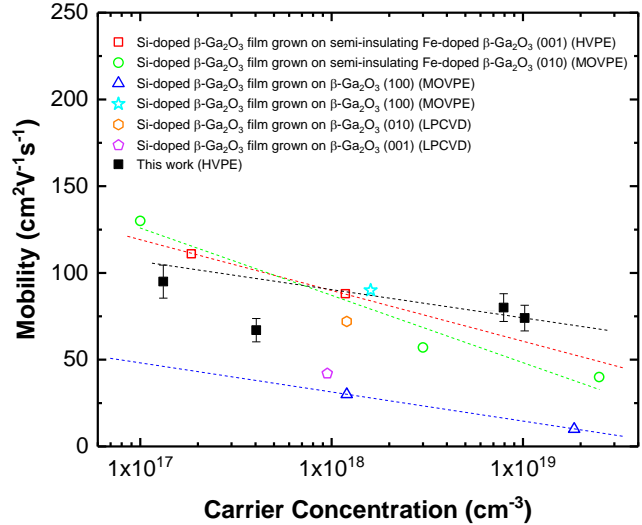

FIG. 2. Electron mobility as a function of carrier concentration at RT for Si doped β-Ga$_2$O$_3$ films homoepitaxially grown by HVPE on semi-insulating Fe doped β-Ga$_2$O$_3$ (010) (n>10$^{18}$cm$^{-3}$) and β-Ga$_2$O$_3$ (010) (n<10$^{18}$cm$^{-3}$) substrates represented by black squares with error bars. The dashed lines are the linear regression of the experimental data, and the open symbols show experimental results from different groups: Ref. 6, □; Ref. 63, ○; Ref. 57, △; Ref. 67, ☆; Ref. 62, ⬡; Ref. 62, ⬠. Primarily, highest reported mobility values for Si doped β-Ga$_2$O$_3$ thin films homoepitaxially grown by various growth methods are aimed to show for comparison.

In order to understand the structural quality of the homoepitaxial films, we carried out RT Raman spectroscopy measurements. These measurements also allow us to compare the role of Si-doping on the phonon vibration modes of β-Ga$_2$O$_3$. The Raman results are summarized in Fig. 3. Since β-Ga$_2$O$_3$ has a monoclinic structure and belongs to the C$_{2h}$ space group[34,68], 11 sharp peaks are observed at 111, 146, 170, 201, 321, 347, 417, 475, 631, 659, 767cm$^{-1}$ for all samples in good agreement with Refs 69, 70, and 71, and with only minor variations for some phonon modes. The Raman spectra for β-Ga$_2$O$_3$ can be classified in three regions[14,42,68,72]: (i) vibration and translation of tetrahedral-octahedral chains Ga$_I$O$_4$ (low frequency, 100-200cm$^{-1}$); (ii) deformation of the tetrahedral and octahedral groups Ga$_I$O$_4$ and Ga$_{II}$O$_6$ (middle frequency, 300-500cm$^{-1}$), and (iii) stretching and bending of tetrahedral groups Ga$_I$O$_4$ (high frequency, 600-800cm$^{-1}$).

In Fig. 3, only phonon modes belonging to β-Ga$_2$O$_3$ are observed. Therefore, we can rule out the presence of other Ga$_2$O$_3$ phases as a consequence of the HVPE growth. In terms of crystalline quality, Figs. 3(a) and 3(b) clearly show that the A$_g$(3) Raman mode has the strongest intensity and its FWHM is the narrowest, which also indicates that Si doped β-Ga$_2$O$_3$ films have high crystalline quality[34]. By comparing the Raman spectra of the Si doped β-Ga$_2$O$_3$ films (Figs. 3(a) and 3(b)) with the β-Ga$_2$O$_3$ (010) substrate (Fig. 3(c)), Si doped β-Ga$_2$O$_3$ films showed similar Raman mode values compared to the β-Ga$_2$O$_3$ (010) substrate, indicating that the Si doped β-Ga$_2$O$_3$ films are under insignificant strain[14,72]. Furthermore, the

$A_g(10)$ Raman mode of the Si doped β-$Ga_2O_3$ films is slightly lower than that of the β-$Ga_2O_3$ substrate, which may be due to the role of impurity doping that results in a weak confined strain in the films.

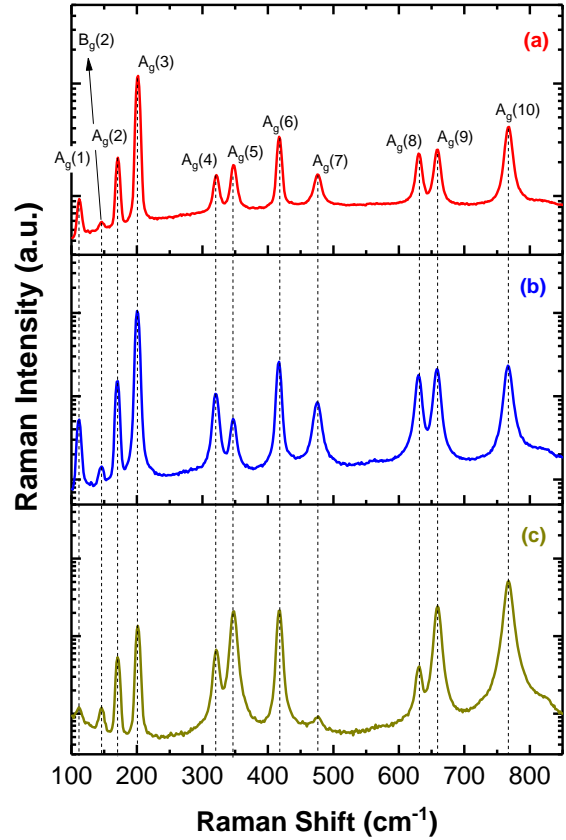

FIG. 3. RT Raman spectra of (a) Si doped β-$Ga_2O_3$ film grown on semi-insulating Fe doped β-$Ga_2O_3$ (010) (b) Si doped β-$Ga_2O_3$ film grown on β-$Ga_2O_3$ (010) (c) β-$Ga_2O_3$ (010) substrate excited at λ=532nm. Y-axis is in logarithmic scale.

The crystalline quality of the Si doped β-$Ga_2O_3$ films was assessed by XRD and GIXRD rocking curve measurements at the Advanced Photon Source. Samples were characterized using various angles of incidence, from ~5° down to ~0.13°. In Figure 4, we show the XRD and GIXRD profiles of the 022 diffraction peak for two for β-$Ga_2O_3$ films grown on Fe doped β-$Ga_2O_3$ (010) [Figs. 4(a) and (b)] and on β-$Ga_2O_3$ (010) [Figs. 4(c) and (d)]. The rocking curves shown in Figs. 4(a) and 4(c) were obtained utilizing an angle of incidence (~5°) that is significantly larger than the critical angle of both the film and the substrate, which results in the X-ray beam penetrating deeply into the substrate. Therefore, the resulting XRD peaks are mainly dominated by scattering from the substrate. In contrast, Figs. 4(b) and 4(d) were obtained using an incident angle (i.e., ~0.13°) that is smaller than the critical angle of the film. This results in total reflection of the X-ray beam, which then penetrate the film only to the few nanometer depth of an evanescent wave. As a result, the resulting peaks observed in Figs. 4(b) and 4(d) are dominated by the film region and have minor contribution from the substrate. By comparing the overall rocking curve FWHMs of the XRD and GIXRD scans, it can be concluded that high crystal quality Si doped β-$Ga_2O_3$ films were realized by homoepitaxially grown on both semi-insulating Fe doped β-$Ga_2O_3$ (010) (Figs. 4(a) and 4(b)) and β-$Ga_2O_3$ (010) (Figs. 4(c) and 4(d)) substrates. However, a comparison of the plots at different incident X-ray beam angles indicate that there is more mosaic in the films than in the substrate.

Furthermore, GIXRD results demonstrate that a Si doped β-$Ga_2O_3$ film grown on Fe doped β-$Ga_2O_3$ (010) (with carrier concentration and electron mobility values of $7.92 \times 10^{18} cm^{-3}$ and $80 cm^2 V^{-1} s^{-1}$, respectively) has a rocking curve FWHM= 0.0361° (Fig. 4(b)). This value is smaller, than that of the Si doped β-Ga$_2$O$_3$ film grown on β-Ga$_2$O$_3$ (010) substrate (with carrier concentration and electron mobility values of 4.04×10$^{17}$cm$^{-3}$ and 67cm$^2$V$^{-1}$s$^{-1}$, respectively), which has a rocking curve FWHM= 0.0965° (Fig. 4(d)). This correlates with the FWHMs of the substrates: the rocking curve FWHM of the Fe doped β-Ga$_2$O$_3$ (010) substrate (Fig. 4(a)) is smaller than the β-Ga$_2$O$_3$ (010) substrate (Fig. 4(c)), which indicates that Fig. 4(a) has likely smaller mosaic than Fig. 4(c). These results also indicate that the β-Ga$_2$O$_3$ grown on the Fe doped β-Ga$_2$O$_3$ (010) has a better crystalline quality despite having an order of magnitude larger carrier concentration. The rapid decrease in electron mobility from n=7.92×10$^{18}$ to 4.04×10$^{17}$cm$^{-3}$ (Fig. 2) associated with a strong deterioration in crystalline quality of the β-Ga$_2$O$_3$ film grown on β-Ga$_2$O$_3$ (010). The PL FWHM values (Table. I) are also in agreement with the GIXRD rocking curve measurements of these two samples. A comprehensive analysis of the synchrotron X-ray diffraction results and their correlation with various defect types such as thread dislocations for the β-Ga$_2$O$_3$ samples will be discussed in a separate paper.

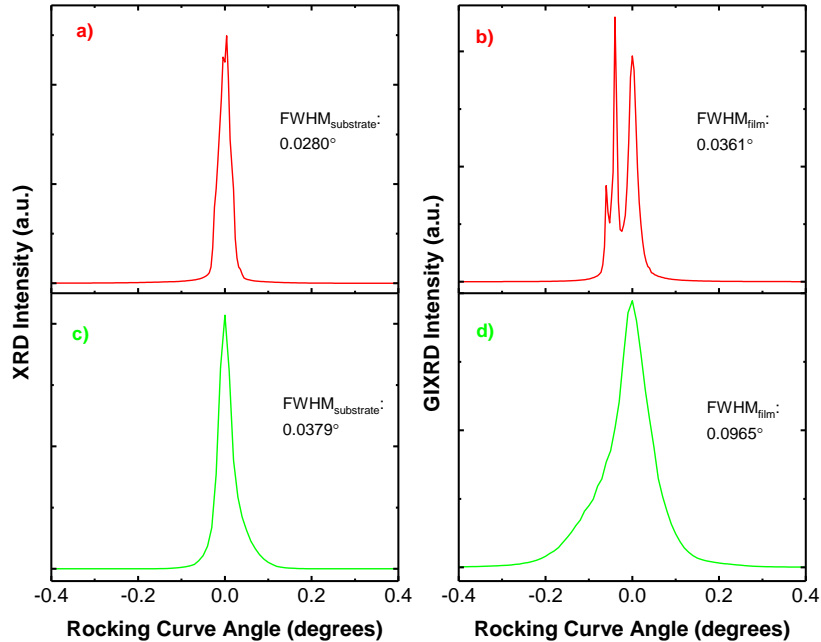

FIG. 4. Rocking curve scans of the 022 reflection of Si doped β-Ga$_2$O$_3$ films homoepitaxially grown by HVPE at high incidence (~5°, left) and grazing incidence (~0.13°, right): (a) and (b) semi-insulating Fe doped β-Ga$_2$O$_3$ (010) substrate (n=7.92×10$^{18}$cm$^{-3}$, μ=80cm$^2$V$^{-1}$s$^{-1}$);(c) and (d) UID β-Ga$_2$O$_3$ (010) (n=4.04×10$^{17}$cm$^{-3}$, μ=67cm$^2$V$^{-1}$s$^{-1}$) substrate.

Several studies have recently reported on power devices of Ga$_2$O$_3$ (e.g., MOSFETs, SBDs, and MOS capacitors), fabricated on β-Ga$_2$O$_3$ in the (010), (100), (001) and (201) crystal orientations to investigate the crystal-orientation related material and device properties. Thin Al$_2$O$_3$, SiO$_2$, and HfO$_2$ films have been widely utilized as gate dielectrics in Ga$_2$O$_3$ power devices owing to their large bandgaps that are required to achieve adequate conduction band offset for Ga$_2$O$_3$[73,74]. Notably, carrier trapping in the gate dielectrics deteriorates device properties such as mobility, gate leakage, etc.[75]. However, considering the thin HfO$_2$ as a gate dielectric, the leakage current is caused by its small conduction band offset with β-Ga$_2$O$_3$[76].

Here, we fabricated a β-Ga$_2$O$_3$ (010) MOS capacitors with HfO$_2$ dielectric layers to analyze their electrical properties. The RT C-V characteristics of 2nm HfO$_2$ from 4V to -4V at 100kHz is shown in Fig. 5(a). The C-V plot noticeably exhibits deep depletion behavior for reversed bias voltages without hysteresis indicating a scarcity of "border" traps in the near interface region of the oxide[74]. The carrier concentration of bulk β-Ga$_2$O$_3$ (010) single crystals were determined by fitting 1/C$^2$-V plots (Fig. 5(b)) to n =

$2/[q\varepsilon_s\varepsilon_0 A^2 d(C^{-2})/dV_g]$[76,77], where q is electron charge, A is anode area, $\varepsilon_0$ is the vacuum permittivity, and $\varepsilon_0 = 10$ is the relative dielectric constant of β-$Ga_2O_3$[78]. From the reciprocal of the slope of the $1/C^2$ vs. V curve, a carrier concentration value of $2.15\times10^{17} cm^{-3}$ was obtained, which is close to the $1.66\times10^{17} cm^{-3}$ value determined via RT hall effect measurements. Additionally, $1/C^2$-V characteristics (Fig. 5(b)) obtained from the data in Fig. 5(a) are linear at voltages from -4 to 0V, which demonstrates a stable carrier density of $2.15\times10^{17} cm^{-3}$ [79]. A non-linear slope over the range from 0 to 1V indicates the presence of residual interface states, as the linearity of the slope is correlated with interface states at the junction interface[80].

The built-in voltage ($V_{bi}$) is 1.67V from a linear extrapolation of the $1/C^2$-V plot, as shown in Fig. 5(b). Furthermore, the dielectric capacitance ($C_{ox}$) is measured as $4.49\times10^{-6} F/cm^2$ (Fig. 5(a)), and a dielectric constant ($\varepsilon_{ox}$) of 10.14 was extracted from the maximum accumulation capacitance for the thin $HfO_2$ film using $C_{ox} = \varepsilon_{ox}\varepsilon_0/t$, where t is the thickness of the dielectric layer. The dielectric constant of the thin $HfO_2$ film is found to be lower than some of the reported dielectric values (≥14) for ALD $HfO_2$[77,79]. The amorphous structure of the $HfO_2$ layer and interlayer roughness between the substrate and the $HfO_2$ can be the reason for the lower dielectric constant[76,79].

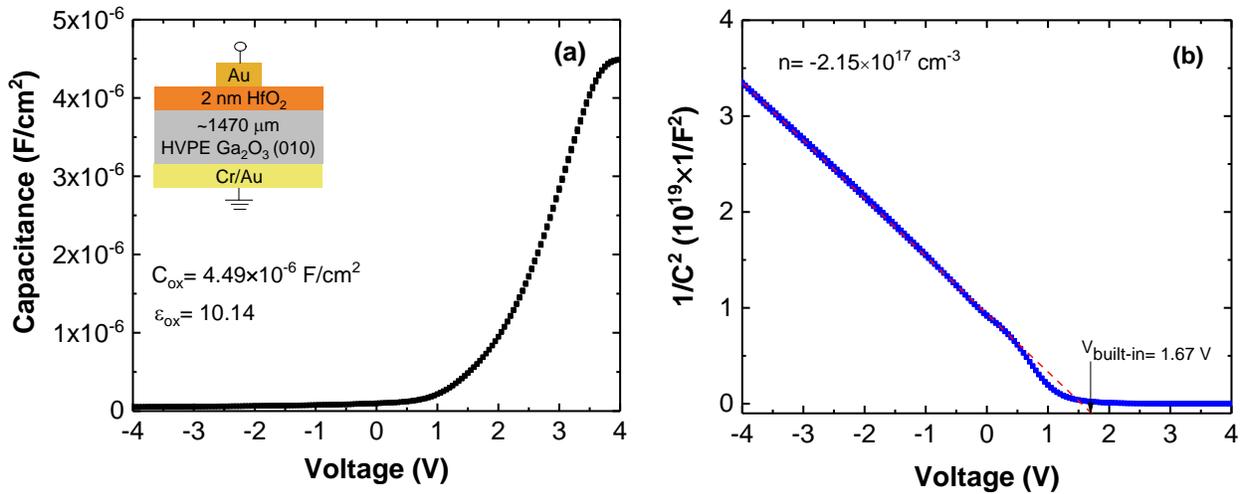

FIG. 5. (a) C-V characteristics for 2nm $HfO_2$ on bulk β-$Ga_2O_3$ (010) single crystal measured at 100kHz. Inset shows the cross-section schematic of the $HfO_2/Ga_2O_3$ MOS capacitor. (b) The $1/C^2$-V plot for C-V sweeps from 4V to -4V, indicating a carrier concentration of $2.15\times10^{17} cm^{-3}$ (close to the carrier concentration of $1.66\times10^{17} cm^{-3}$ obtained via RT hall effect measurement). The built-in voltage ($V_{bi}$) is extracted from the X-axis intercept of the extrapolation of the linear part of the data, and the bulk β-$Ga_2O_3$ (010) single crystal doping concentration n is given by the slope of the linear fit.

### 4. Conclusions

In summary, we have investigated the RT optical, structural and electrical properties of Si doped β-$Ga_2O_3$ films grown on semi-insulating Fe doped and bulk β-$Ga_2O_3$ (010) substrates by HVPE. PL measurements revealed a strong, broad emission peak around 3.34eV in the UV region, which is attributed to an intrinsic transition between $V_o$ and $N_{int}$, owing to the recombination of a self-trapped exciton. Less intense broadband peaks also emerged in the range of 1.74-1.76eV, which can result from the recombination of the deeply trapped exciton.

Furthermore, the Raman spectra demonstrated only β-$Ga_2O_3$ phonon modes, which excludes the existence of other phases and indicates that the β-$Ga_2O_3$ thin films are pure and exhibits good crystal quality. The synchrotron GIXRD measurements revealed that despite the higher growth rates afforded by HVPE, β-$Ga_2O_3$ thin films grown on Fe doped β-$Ga_2O_3$ (010) and β-$Ga_2O_3$ (010) substrates have high crystal quality. Si doping of the β-$Ga_2O_3$ films achieved carrier concentrations in the range of ~$10^{17}$-$10^{19} cm^{-3}$, and a mobility value of 95cm$^2$V$^{-1}$s$^{-1}$ for a carrier concentration of 1.32×10$^{17}$cm$^{-3}$. For the carrier concentrations above 10$^{18}$cm$^{-3}$, the mobility values of the Si doped β-Ga$_2$O$_3$ films E showed a higher trend compare to the reported literature values grown by other techniques. Our results also demonstrate that Si is an effective n-type dopant for HVPE-grown β-Ga$_2$O$_3$ films, yielding higher carrier concentrations without sacrificing electron mobility compared to other dopants. Moreover, RT C-V measurements of MOS capacitors fabricated using the HVPE-grown bulk β-Ga$_2$O$_3$ (010) single crystal yielded carrier concentration values similar to the values derived from Hall effect measurements.

The comparison between high and grazing incidence XRD indicates that, despite generally larger growth rates, HVPE can lead to high quality epitaxial films, albeit with a higher mosaic spread than those of the bulk. The difference between XRD and GIXRD results emphasizes the need to reach incidence angles below the critical angle to probe the epitaxial quality of β-Ga$_2$O$_3$. More systematic studies are needed though to understand the origin of the different contributions, as well as the impact that growth conditions can have on the overall epitaxial quality of the films.

These results suggest that HVPE is a promising method for growing high-quality Si doped β-Ga$_2$O$_3$ films on bulk β-Ga$_2$O$_3$ single crystal (010) substrates. In addition, this work provides a further understanding of the fundamental properties of β-Ga$_2$O$_3$ films homoepitaxially grown by HVPE, which should enable the development of high-performance electronic devices.

**Acknowledgements**

The authors would like to thank Liliana Stan for fabrication of the metal contacts for electrical measurements and Dr. Sushant Sonde for the fruitful discussion and generous support for device fabrication. Argonne National Laboratory's contribution is based upon work supported by Laboratory Directed Research and Development (LDRD) funding from Argonne National Laboratory, provided by the Director, Office of Science, of the U.S. Department of Energy under Contract No. DE-AC02-06CH11357. The authors would also like to acknowledge the U.S. Department of Energy's Technologist in Residence program for facilitating the collaboration between Argonne and Kyma. This research used the resources of the Center for Nanoscale Materials, an Office of Science user facility supported by the U.S. Department of Energy, Office of Science, Office of Basic Energy Sciences, and the Advanced Photon Source, also supported by the U. S. Department of Energy, Office of Science, Office of Basic Energy Sciences, under Contract No. DE-AC02-06CH11357.